\documentclass[
superscriptaddress,
preprint,
nofootinbib,
 amsmath,amssymb,
 aps,
prd,longbibliography,
]{revtex4-2}
\usepackage{multirow}
\usepackage{xcolor}
\usepackage{graphicx}
\usepackage{dcolumn}
\usepackage{bm}
\usepackage{hyperref}

\newcommand{\be}{\begin{equation}}
\newcommand{\ee}{\end{equation}}
\newcommand{\ba}{\begin{align}}
\newcommand{\ea}{\end{align}}
\newcommand{\half}{\frac{1}{2}}
\newcommand{\bi}{\begin{itemize}}
\newcommand{\ei}{\end{itemize}}

\let\a=\alpha \let\b=\beta \let\g=\gamma \let\d=\delta \let\e=\epsilon
\let\z=\zeta    \let\k=\kappa
\let\l=\lambda \let\m=\mu \let\n=\nu  \let\p=\pi \let\r=\rho
\let\s=\sigma \let\t=\tau  \let\f=\phi \let\c=\chi \let\ps=\psi
\let\w=\omega    \let\D=\Delta  \let\L=\Lambda
  \let\S=\Sigma   
 
\let\we=\wedge
\let\ph=\phantom
\let\pt=\partial

\makeatletter
\newcommand*{\Rom}[1]{\expandafter\@slowromancap\romannumeral #1@}
\makeatother

\makeatletter
\newcommand*{\rom}[1]{\expandafter\romannumeral #1}
\makeatother

\let\nn=\nonumber

\def\beq#1\eeq{\begin{align}#1\end{align}}

\begin{document}


\title{Wrapped Brane Solutions in Romans $F(4)$ Gauged Supergravity}

\author{Nakwoo Kim}
 \email{nkim@khu.ac.kr}

\affiliation{Department of Physics and Research Institute of Basic Science, Kyung Hee University, Seoul 02447, Korea}
 \affiliation{School of Physics, Korea Institute for Advanced Study, Seoul 02445, Korea}

\author{Myungbo Shim}%
 \email{mbshim1213@khu.ac.kr}
 \affiliation{Department of Physics and Research Institute of Basic Science, Kyung Hee University, Seoul 02447, Korea}

\begin{abstract}
We explore the spectrum of lower-dimensional anti-de Sitter (AdS) solutions in $F(4)$ gauged supergravity in six dimensions. The ansatz employed corresponds to D4-branes partially wrapped on various supersymmetric cycles in special holonomy manifolds. Re-visiting and extending previous results, we study the cases of two, three, and four-dimensional supersymmetric cycles within Calabi-Yau threefold, fourfold, $G_2$, and Spin(7) holonomy manifolds. We also report on non-supersymmetric AdS vacua, and check their stability in the consistently truncated lower-dimensional effective action, using the Breitenlohner-Freedman bound. We also analyze the IR behavior and discuss the admissibility of singular flows.
\end{abstract}

\maketitle
\newpage
\section{\label{sec:level1}Introduction}
Exploring the spectrum of supersymmetric anti-de Sitter solutions in String/M-theory is an intriguing enterprise, not only for the beauty of its mathematical structure  but also due to AdS/CFT correspondence \cite{Maldacena1999g}. Maximally supersymmetric solutions, AdS${}_4\times S^7$, AdS${}_5\times S^5$, and AdS${}_7\times S^4$ are well-known, but we are eventually interested in the duality of more realistic gauge field theories, so constructing less-supersymmetric AdS backgrounds which do have 10 or 11 dimensional supergravity origin is a valuable endeavor. For this purpose one usually takes one of the following two approaches in search of supersymmetric AdS solutions. The first approach is to study the most general form of supersymmetric AdS solutions in the dimensions of interest, using the geometry of Killing spinors. One sometimes manages to find new solutions \cite{Gauntlett:2002nw,Gauntlett:2004yd}, or discover interesting novel geometric structures {\it e.g.} in \cite{Kim:2005ez,Kim:2006qu,Gauntlett:2006ns,Gauntlett:2007ts,Donos:2008ug}. On the other hand, one can utilize lower-dimensional gauged supergravity models which are consistent truncation of 10 or 11 dimensional supergravities. Interesting AdS solutions may be obtained by studying the critical points of the scalar potential or considering spontaneous dimensional reduction by turning on various gauge fields, see {\it e.g.} \cite{Pernici:1984nw}. 

In the latter construction, supersymmetry is preserved if one turns on magnetic fields whose values are the same as curvature tensor on twisted part of the space, and in string theory interpretation the solutions correspond to branes partially wrapped on supersymmetric cycles of various dimensionalities \cite{Maldacena2001f,Maldacena2001g,Acharya2001,Gauntlett2001b,Gauntlett2001,Nunez2001,Gauntlett2002c,Naka2002}. See also {\it e.g.} \cite{Gauntlett2007a} and \cite{Kim2014}, for non-AdS configurations and generalizations. Note that to be precise one additionally imposes Killing spinor projection rules so that the effect of spin connection and gauge connection cancel along the ``supersymmetric cycle'' directions. The solutions, in particular the curvature radius of AdS and the supersymmetric cycle part, must contain information on D- or M-brane worldvolume theory with topological twisting \cite{Maldacena2001g}.

Back to the long list of AdS vacua in String/M-theory, precise matching with large-$N$ limit of dual quantum field theory has been missing for a long time, except for ${\cal N}=4$ super Yang-Mills which can be probed perturbatively. However, 
thanks to the localization technique \cite{Pestun:2007rz,Kapustin:2009kz}, and also to the realization of M2-brane dynamics as Chern-Simons matter system \cite{Aharony2008}, it is now possible at least for favorable backgrounds to match quantitatively various physical quantities between the two sides of the duality relation, see for AdS${}_4$/CFT${}_3$ examples {\it e.g.} \cite{Drukker:2010nc,Herzog2011,Martelli:2011qj,Cheon:2011vi,Jafferis2011}. It is, of course, possible to extend this prediction to the case of topologically twisted supersymmetric field theory obtained from branes wrapped on supersymmetric cycles. For a particular class of such relations, namely M5-branes wrapped on 3-cycles and their description as Chern-Simons theory, see {\it e.g.} \cite{Gang2014,Gang2015,Gang2016,Gang2016a,Gang2019,Gang:2019uay}.

In this article, we are interested in wrapped brane configurations which lead to two, three, and four-dimensional AdS vacua, in $D=6$ $F(4)$ gauged supergravity \cite{Romans1986d}. It has long been known that this particular theory is a consistent truncation of 
$D=10$ massive type IIA supergravity \cite{Cvetic1999}. Note that it is also established more recently that this theory can be uplifted to IIB supergravity as well \cite{Jeong2013b,Hong2018b,Malek2019}. For definiteness we will consider in this article uplifts to massive IIA\footnote{In IIB setting we have a network of D5 and NS5-branes preserving $(4+1)$-dimensional Lorentz symmetry.}, where the relevant brane interpretation is as D4-branes in the presence of D8-branes. The AdS vacuum of $F(4)$ gauged supergravity has 16 supercharges, and the dual field theory is proposed to be a five-dimensional supersymmetric gauge theory with $USp(2N)$ gauge group and $N_f<8$ massless hypermultiplets in fundamental representation \cite{Seiberg:1996bd,Morrison:1996xf,Intriligator:1997pq,Brandhuber1999}, and see also \cite{DAuria2000,DAuria2001a,Imamura2001,Karndumri2013,Karndumri2014,Chang2018a,Dibitetto2019} for gravity side analysis. The duality was checked using localization formula and the results for entanglement entropy agree with $N^{5/2}$ scaling of degrees of freedom \cite{Jafferis2014}.

Among the AdS solutions from wrapped branes, AdS${}_2$ solutions can be interpreted as near horizon limit of magnetically charged black holes, and, on the field theory side, the entropy is associated with the topologically twisted index. For M2-brane theory, agreement between the two sides of AdS/CFT was shown in \cite{Benini2015,Benini2016}. As one tries to apply this relation to black holes in $F(4)$ gauged supergravity, the field theory computations in \cite{Hosseini2018a,Hosseini:2018usu,Crichigno2018} and the supergravity side result match \cite{Suh2019,Suh2018a,Suh2018}, \emph{only after} a mistake in \cite{Naka2002} is fixed: in this reference, an instanton-like contribution for four-cycles was overlooked, and a correct solution for K\"ahler 4-cycle was presented by M. Suh \cite{Suh2019}. This realization has prompted our present work. We re-visit the construction of AdS solutions from wrapped branes in $F(4)$ gauged supergravity, and provide a list of supersymmetric and non-supersymmetric solutions. We fill other gaps in \cite{Naka2002} by studying also the flows between AdS${}_6$ and configurations with lower-dimensional Lorentz symmetry, and study the admissibility IR singularities following the criteria of Maldacena-Nu\~nez \cite{Maldacena2001g} and Gubser \cite{Gubser2000b}. We also provide the consistently truncated lower-dimensional actions, in the manner of \cite{Gauntlett2002}, and study the fluctuation modes to see if they violate the Breitenlohner-Freedman bound \cite{Breitenlohner:1982jf} for stability. For some of recent works on holography of $F(4)$ gauged supergravity, readers are referred also to \cite{Alday2014,Gutperle2017a,Malek2018}.

The plan of this article is as follows. In Sec.\ref{sec:2}, we present the action of $F(4)$ gauged supergravity and the BPS conditions along with the BPS equations for wrapped brane configurations. In Sec.\ref{sec:3}, we present numerical solutions for holographic RG flows from the AdS${}_6$ in UV. Apart from the special solutions which flow exactly to lower-dimensional AdS solutions in IR, the solutions are generically singular in IR, and we check if those singularities are physically allowed, employing the criteria of Refs. \cite{Maldacena2001g} and \cite{Gubser2000b}. In Sec.\ref{sec:4}, we present lower-dimensional consistently truncated action, look for non-supersymmetric AdS fixed points, and study their stability using the Breitenlohner-Freedman bound. We conclude in Sec.\ref{sec:5}, and in the appendices readers may find some of detailed computations the result of which are presented in the main text. 
\section{F(4) Gauged Supergravity}
\label{sec:2}
\subsection{The Action and Its Relation to 10 Dimensions}
Let us first start by presenting the action of the bosonic sector for $D=6$, $F(4)$ gauged supergravity.
\begin{align}
\mathcal{S}_{F(4)}=&\frac{1}{2\k_{6}^{2}}\int d^{6}x\sqrt{-g}\ph{.}[\frac{1}{4}R-\frac{1}{2}\pt_{\m}\f\pt^{\m}\f+\frac{1}{8}(g^2e^{\sqrt{2}\f}+4gme^{-\sqrt{2}\f}-m^2e^{-3\sqrt{2}\f}) \nn\\
&-\frac{1}{4}e^{-\sqrt{2}\f}(\mathcal{H}_{\m\n}\mathcal{H}^{\mu\nu}+F^I_{\mu\nu}F^{I\mu\nu})-\frac{1}{12}e^{2\sqrt{2}\f}G_{\m\n\r}G^{\m\n\r} \nn \\
&-\frac{1}{8}\e^{\m\n\r\s\t\k}B_{\m\n}(\mathcal{F}_{\r\s}\mathcal{F}_{\t\k}+mB_{\r\s}\mathcal{F}_{\t\k}+\frac{1}{3}m^2B_{\r\s}B_{\t\k}+F^I_{\r\s}F^I_{\t\k})]  .
\end{align}

The action as it stands includes gravity via metric $g_{\mu\nu}$, a two-form tensor field $B$ with field strength $G=dB$, a triplet of $SU(2)$ gauge fields $A^I$, a $U(1)$ vector $\mathcal{A}$, and a real scalar field $\phi$. $\mathcal{H}$ is a combination of the field strength $\mathcal{F}=d\mathcal{A}$ and two-form tensor field, namely
$\mathcal{H}=\mathcal{F}+mB$. Note that the total number of on-shell bosonic degrees of freedom is $32$.
We have two coupling constants, $g$ and $m$, in addition to Newton constant $\kappa_6$. 

When we uplift this system to $D=10$ massive IIA supergravity, we have $m_{10d}=\sqrt{2}m$ in the convention of {\it e.g.} \cite{Cvetic1999,Naka2002}. We note that 
there exist alternative embeddings into type \Rom{2}B theory which was recently found in \cite{Jeong2013b,Hong2018b}\footnote{The associated supersymmetric AdS${}_6$ solutions in \Rom{2}B are studied in {\it e.g.} \cite{Apruzzi:2014qva,Kim2015,Kim2016,DHoker2016,DHoker2017,DHoker2017a,DHoker2017b}. It is also established that IIA and IIB descriptions are related through a non-Abelian T-duality \cite{Sfetsos:2010uq,Itsios:2013wd,Kelekci2015,Terrisse:2018hhf,Lozano:2018pcp}.}, and also into the exceptional field theory formalism \cite{Malek2018, Malek2019}. This theory allows a supersymmetric AdS${}_6$ solution when 
$e^{-2\sqrt{2}\f}=g/3m$ and all other bosonic (and also fermionic) fields are trivial. When uplifted, it is a 1/2-BPS configuration of IIA/IIB supergravity in $D=10$. In the convention we adopt, the radius of AdS space is $L_{AdS_{6}}=3\sqrt{2}(3mg^{3})^{-1/4}$, or $3\sqrt{2}/g$ when we substitute $m=g/3$ as a convenient choice for the theory which can be uplifted in IIA/IIB.

Let us record here that convention we adopt is related to the one in \cite{Cvetic1999} in the following way.\footnote{Note that $g=3m$ is not essential for embedding the theory in $D=10$ supergravity.}
\ba
g_{there}=\half g,&&m_{there}=\half m_{10d},
\end{align}
\ba
A^{(p)}_{there}=2A^{(p)},&&X_{there}=e^{-\frac{1}{2\sqrt{2}}\f_{there}}=e^{\frac{1}{\sqrt{2}}\f},&&\mathcal{L}_{there}=4\mathcal{L},
\end{align}
where $A^{(p)}$ is a $p$-form potential in $D=6$ theory. 

\subsection{A Survey of Wrapped Brane Solutions}
In this paper, we are interested in a specific type of classical solutions: in particular, we have lower-dimensional anti-de Sitter spaces in mind. This type of solutions were known as ``magnetovacs'' before the advent of string duality and D-branes \cite{Pernici:1984nw}. Thanks to a seminal paper of Maldacena-Nu\~nez \cite{Maldacena2001f}, and the extension to higher-dimensional cycles \cite{Acharya2001,Gauntlett2001b,Gauntlett2002c}, these solutions are nowadays commonly referred to ``wrapped-brane'' solutions. The way how to produce such non-trivial solutions is as follows. We assume that part of the space is Einstein (which corresponds to supersymmetric cycles), and turn on gauge connection and impose Killing spinor projection rules so that the contributions of spin connection and gauge connection exactly cancel, at least along the cycle directions. This is the manifestation of topological twisting (via cancelling the spin connection, we effectively turn a spinor into a scalar). Depending on the concrete choice of gauge connection, we deal with different kinds of special holonomy manifolds and supersymmetric cycles thereof. 

More concretely, the metric ansatz goes like 
\be
ds^{2}_6=e^{2f(r)}(-dt^{2}+dr^{2}+\sum_{\a=1}^{4-d}dx^{2}_{\a})+\sum_{i}e^{2\l_{i}(r)}ds_{\mathcal{M}_{i,d}}^{2} \, . 
\ee
On the right-hand-side, the part with scale factor $e^{2f}$ contains the reduced worldvolume (after wrapping) and the ``holographic'' coordinate $r$. Then the latter part with scale factors $e^{2\l_i}$ denotes the ``supersymmetric cycle''. For our purposes here, this part is either a single Einstein space or a sum of two Einstein spaces up to scale factors which is a function of $r$ only. The cycle part will be chosen as (sum of) constant curvature spaces, {\it e.g.} the sphere $S^d$, the complex projective manifold $\mathbb{CP}^n$, and their negatively-curved cousins such as the hyperbolic manifold and the Bergman space for concreteness. 
Then we turn on magnetic field for the $SU(2)$ part of the vector fields. The point is to make sure the effect of spin connection and gauge connection cancel along the cycle directions for spinors satisfying certain projection rules. 

The resulting BPS equations are always given in the following way.
\ba
f'e^{-f}&=-\frac{1}{4\sqrt{2}}\left[ge^{\frac{1}{\sqrt{2}}\f}+me^{-\frac{3}{\sqrt{2}}\f}-\sum_{i}\frac{\D_{i} k_{i}}{g} e^{-\frac{1}{\sqrt{2}}\f-2\l_{i}(r)}\right]+3\Upsilon e^{\frac{1}{\sqrt{2}}\f-\sum_{i}\D_{i}\l_{i}(r)},\nn
\\
\l_{i}'e^{-f}&=-\frac{1}{4\sqrt{2}}\left[ge^{\frac{1}{\sqrt{2}}\f}+me^{-\frac{3}{\sqrt{2}}\f}+\sum_{j}\frac{\tilde{\D}^{(i)}_{j}k_{j}}{g} e^{-\frac{1}{\sqrt{2}}\f-2\l_{j}(r)}\right]-\Upsilon e^{\frac{1}{\sqrt{2}}\f-\sum_{j}\D_{j}\l_{j}(r)},
\\
\frac{\f'}{\sqrt{2}}e^{-f}&=-\frac{1}{4\sqrt{2}}\left[-ge^{\frac{1}{\sqrt{2}}\f}+3me^{-\frac{3}{\sqrt{2}}\f}+\sum_{i}\frac{\D_{i}k_{i}}{g} e^{-\frac{1}{\sqrt{2}}\f-2\l_{i}(r)}\right]+\Upsilon e^{\frac{1}{\sqrt{2}}\f-\sum_{i}\D_{i}\l_{i}(r)},\nn
\end{align}
where $\sum_{i}\D_{i}=d$, $\sum_{j}\tilde{\D}^{(i)}_{j}=(8-d)$, and $\Upsilon$ is zero for $d=2,3$, while non-zero for $d=4$. Fixed points, {\it i.e.} lower-dimensional AdS spaces, arise when the radii of cycles $\lambda_i$ and the scalar $\phi$ are constants. We have checked that all solutions in previous works  \cite{Naka2002,Suh2019}\footnote{The correct solutions for 2-, 3-cycles and the solutions for 4-cycles are, respectively, discovered in the former and the latter.} can be obtained from these BPS equations.

Explicit ansatz for each case is summarized in Table \ref{tab:table1}, and the properties of the solutions are provided in Table \ref{tab:table2}.

\begin{table*}
\begin{ruledtabular}
\begin{tabular}{cccc}
 Cycles&$\mathcal{F}$&$F^{\hat{I}}_{\m\n}$&$B_{\m\n}$
\\ \hline
 2-Cycles&$0$&$F^{\hat{3}}_{45}=\frac{k\z}{g}e^{-2\l}$ &0
 \\
 3-Cycles&0 &$F^{\hat{I}}_{\textrm{non-zero}}=\frac{k\z_{I}}{2g}e^{-2\l}$&0
 \\[0.3em]
 
Cayley 4-Cycles &0&$F^{\hat{I}}_{\textrm{non-zero}}=\frac{k\z_{I}}{3g}e^{-2\l}$&$B_{01}=-\frac{2}{3m^{2}g^{2}}e^{\sqrt{2}\f-4\l}$
  \\[0.3em]
  
 K\"ahler 4-Cycles&0&$F^{\hat{3}}_{23}=F^{\hat{3}}_{45}=\frac{k\z}{g}e^{-2\l}$&$B_{01}=-\frac{2}{m^{2}g^{2}}e^{\sqrt{2}\f-4\l}$
  \\
 K\"ahler $\S_{\mathfrak{g}_{1}}\times\S_{\mathfrak{g}_{2}}$&0&$F^{\hat{3}}_{23}=\frac{k_{1}\z}{g}e^{-2\l_{1}}$, $F^{\hat{3}}_{45}=\frac{k_{2}\z}{g}e^{-2\l_{2}}$
 &$B_{01}=-2\frac{k_{1}k_{2}}{m^{2}g^{2}}e^{\sqrt{2}\f-2(\l_{1}+\l_{2})}$
\end{tabular}
\end{ruledtabular}
    \caption{\label{tab:table1}The ansatz for gauge fields in orthonormal bases for each case. Non-vanishing components are easily read off from the twisting condition. $\z_{(I)}$ is $\pm1$, representing the choice of orientation of wrapped branes. It is also constrained by $\z_{1}\z_{2}\z_{3}=1$. $k=\pm 1$ gives the sign of scalar curvature of the supersymmetric cycles. }
\end{table*}

\begin{table*}
\begin{ruledtabular}
\begin{tabular}{ccccc}
 \multirow{2}{*}{Cycles}&\multirow{2}{*}{$k$}&BPS&Non-BPS&Does non-BPS solution \\
 &&solution&solution& violate the BF Bound? \\
 \hline
  \multirow{2}{*}{2-Cycles}&$1$&X&X&-
 \\
&$-1$ &O &O & Yes
 \\
\hline\multirow{2}{*}{3-Cycles}&$1$&X&X&-
 \\
 &$-1$ &O &O & Yes
\\
\hline $\mathbb{H}_{2}\times\mathbb{H}_{2}$&$(-1,-1)$&O &X&-
 \\
\hline $S^{2}\times S^{2}$&$(1,1)$&X&O&No
 \\
 \hline$S^{2}\times\mathbb{H}_{2}$&$(1,-1)$&X&X&-
 \\\hline
\multirow{2}{*}{K\"ahler 4-Cycles}&1&X&O&No
 \\
 &$-1$&O &X &- 
 \\\hline
\multirow{2}{*}{Cayley 4-Cycles} &1&X&X&-
\\
 &$-1$&O &X &-
 \\
\end{tabular}
\end{ruledtabular}
\caption{\label{tab:table2}A summary of existence of wrapped brane solutions in $F(4)$ gauged supergravity. 
}
\end{table*}
\section{Holographic RG Flows for Supersymmetric Solutions}
\label{sec:3}
\subsection{2 and 3 Cycles}
Let us start with the cases of 2- and 3-cycles. They are relatively simple since the tensor field vanishes, so we treat them collectively. For the former the $SO(2)$ spin connection is identified with $U(1)\subset SU(2)$, and for the latter we identify $SO(3)$ spin connection with the entire $SU(2)$ gauge connection. In IIA description, 2-cycle is inside Calabi-Yau threefold, and for the latter we have associative 3-cycles inside $G_2$ holonomy manifold. 
The setup for vector fields and the projection rule is as given in the table below. Note that except for AdS fixed point solutions, we have an additional projection involving the radial direction. However, AdS fixed point solutions do not require such an extra condition, hence exhibit supersymmetry enhancement. Here $T^{\hat i}=i\sigma^i/2$ is $SU(2)$ generator (anti-Hermitian), and $\gamma^a$ are $D=6$ gamma matrices. Parameters $\z,\z_i$ are $\pm 1$ and represent the choice of BPS conditions, and in particular $\z_1\z_2\z_3=1$.\footnote{We point out that it is obviously inconsistent to set $\z_1=\z_2=\z_3$, and correct an error in eq.(4.13) of Ref.\cite{Suh2019}.}
\begin{center}
\begin{tabular}{|c|cc|}
\hline
2-cycles  &$\w_{45} =  \z gA^{\hat{3}}$,&$\ph{a} T^{\hat{3}}\e=-\half\z\g^{45}\e$
\\ \hline
 &$\w_{34} =  \z_1 gA^{\hat{1}}$,&$\ph{a} T^{\hat{1}}\e=-\half\z_1 \g^{34}\e$
\\
3-cycles &$\w_{53} = \z_2 gA^{\hat{2}}$,&$\ph{a} T^{\hat{2}}\e=-\half\z_2\g^{53}\e$
\\
&$\w_{45} =  \z_3 gA^{\hat{3}}$,&$\ph{a} T^{\hat{3}}\e=-\half\z_3\g^{45}\e$
\\
\hline
\end{tabular}
\label{t2}
\end{center}

We set the tensor field to zero, and there are three functions we need to determine: $f,\lambda,\phi$. 
BPS equations are given below, where $d=2,3$ and denote the dimensionality of the supersymmetric cycles. Note that $k=1$ is for the sphere and $k=-1$ is for the hyperbolic spaces, with constant curvature.

\ba
f'e^{-f}&=-\frac{1}{4\sqrt{2}}\left[ge^{\frac{1}{\sqrt{2}}\f}+me^{-\frac{3}{\sqrt{2}}\f}-\frac{dk}{g} e^{-\frac{1}{\sqrt{2}}\f-2\l(r)}\right],\nn
\\
\l'e^{-f}&=-\frac{1}{4\sqrt{2}}\left[ge^{\frac{1}{\sqrt{2}}\f}+me^{-\frac{3}{\sqrt{2}}\f}+\frac{(8-d)k}{g} e^{-\frac{1}{\sqrt{2}}\f-2\l(r)}\right],\nn
\\
\frac{\f'}{\sqrt{2}}e^{-f}&=-\frac{1}{4\sqrt{2}}\left[-ge^{\frac{1}{\sqrt{2}}\f}+3me^{-\frac{3}{\sqrt{2}}\f}+\frac{dk}{g} e^{-\frac{1}{\sqrt{2}}\f-2\l(r)}\right].
\end{align}

To facilitate the analysis, we find it convenient to introduce new variables as follows, as advocated in \cite{Gauntlett2001b}.
The BPS equations take a bit simpler form in terms of  $x:=e^{2\l-\sqrt{2}\f}$ and $F:=xe^{2\sqrt{2}\f}$,
\ba
\frac{dF}{dx}&=\frac{2F[2k+mgx]}{x(g^{2}F-mgx+(4-d)k)}.
\label{F23}
\end{align}
We find there exist AdS fixed points for $k=-1$:
\ba
d=2: \quad F=4/g^{2},\quad x=2/gm, \\
d=3: \quad F=3/g^{2},\quad x=2/gm.  
\end{align}

\begin{figure}
    \centering
    \includegraphics[scale=0.7]{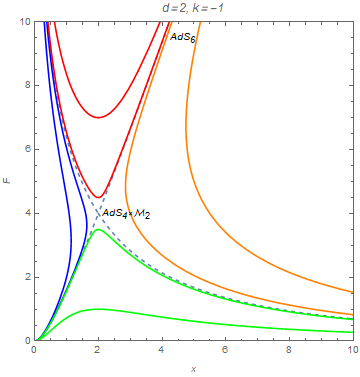}\includegraphics[scale=0.7]{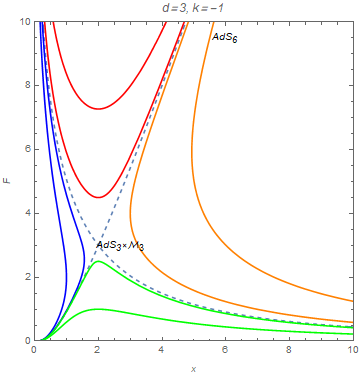}
    \caption{Flow Diagrams for negatively curved 2 and 3 cycles ($k=-1$)}
    \label{fig:1}
\end{figure}

We have not managed to integrate Eq.\eqref{F23} explicitly. However,
in UV regime where $g_{tt}$ becomes large and the metric asymptotes to AdS${}_6$, one easily sees that the solution can be written in a series expansion form,
\be\label{eq:8}
F= \frac{3m}{g}x+\frac{3dk}{g^{2}}+\frac{1}{g^2}\sum_{n=1}^{\infty}\frac{c_{n}}{(mgx)^{\frac{n}{2}}}.
\ee
Here $c_1$ is an integration constant which parametrizes different solutions, and $c_n \,(n>1)$ can be determined recursively in terms of $c_1$. Numerically we find that the flows to $AdS_4\, (d=2)$ and $AdS_3\, (d=3)$ correspond to $c_1=9.1296$ and $c_1={13.951}$ respectively.

\begin{figure}
    \centering
    \includegraphics[scale=0.7]{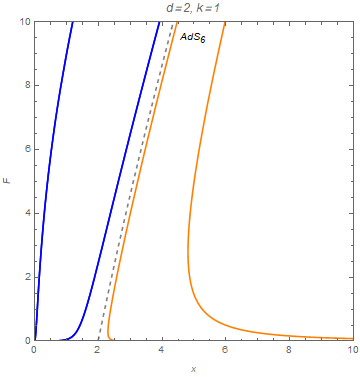}\includegraphics[scale=0.7]{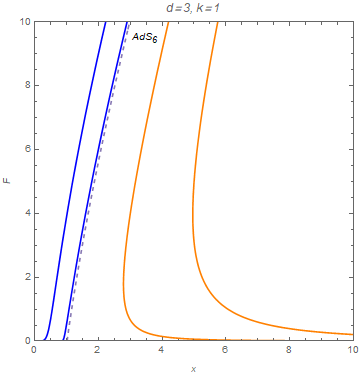}
    \caption{Flow Diagrams for positively curved 2 and 3 cycles, $k=1$}
    \label{fig:2}
\end{figure}
Other than the flows to AdS fixed points, 
there are three different kinds of ``IR'' singularities, according to Figure \ref{fig:1}. 

It turns out that all the singularities are good under the criterion of Ref.\cite{Maldacena2001g}, which instructs us to study the behavior of $g_{tt}$. On the other hand, under the criterion of Ref.\cite{Gubser2000b}, where it was suggested we check the behavior of scalar potential, the singularities with small $F$ are bad. It turns out that the latter criterion is more strict for the solutions at hand. 
When $k=1$, there is no fixed point, but there are flows to good IR singularities (Fig.2).

Below is a summary of the analyses on the type of singularities. We see that the classification is not clear-cut, in particular for solutions with $F\rightarrow 0$ and $x\rightarrow\infty$.
\begin{center}   
\begin{tabular}{|c|c|c|c|c|c|c|}
\hline
 $k$&$x$&$F$&$e^{2f}$&$\left|g_{tt}^{10d}\right|$&$V(\f)$&Type\\ \hline
  $\pm1$&$\infty$&0&0&0&$\infty$(bad)&-
 \\
$\pm1$&0&0&$\infty$&$\infty$ &$\infty$(bad)&Bad
 \\
$-1$&0&$\infty$&0&0&$-\infty$&Good
\\\hline
\end{tabular}
\end{center}
\subsection{4-Cycles}
Now let us turn to 4-cycles. There are two choices for partial twisting now: one is K\"ahler 4-cycle inside a Calabi-Yau 4-manifold, and the other is Cayley 4-cycle inside Spin(7) holonomy manifold. For the former, we identify the $U(1)$ part of spin connection with $U(1)\subset SU(2)$ part of the gauge connections. And for the latter, we set the $SU(2)$ gauge fields to the self-dual part of the spin connection. The BPS conditions are given as below. 
\begin{center}
\begin{tabular}{|c|cc|}
\hline
K\"ahler 4-cycle    & $\w_{23}\pm\w_{45}=g\z A^{\hat{3}}$,& $\half\g_{23}\e=\pm\half\g_{45}\e=-\z T^{\hat{3}}\e$ \\\hline
Cayley cycle      & $\w_{23}\pm\w_{45}=g\z_1 A^{\hat{1}}$,& $\half\g_{23}\e=\pm\half\g_{45}\e=-\z_1 T^{\hat{1}}\e$
\\
$\g^{\mp}_{ij}\e=0$, &$\w_{42}\pm\w_{35}=g\z_2 A^{\hat{2}}$,& $\half\g_{42}\e=\pm\half\g_{35}\e=-\z_2 T^{\hat{2}}\e$
\\
($i,j=2,\cdots,5$)&$\w_{34}\pm\w_{52}=g\z_3 A^{\hat{3}}$,& $\half\g_{34}\e=\pm\half\g_{52}\e=-\z_3 T^{\hat{3}}\e$\\\hline
\end{tabular}
\end{center}
 In the above $\z,\z_i$ are $\pm1$, and $\z_i$ are constrained by $\z_1\z_2\z_3=1$. The associated BPS equations are presented below, where a constant 
$\Upsilon$ denotes non-vanishing instanton density and takes different values for K\"ahler $(\Upsilon=-\frac{1}{\sqrt{2}g^{2}m})$ and Cayley 4-cycles $(\Upsilon=-\frac{1}{3\sqrt{2}g^{2}m})$.
\ba
\label{bps4c}
f'e^{-f}&=-\frac{1}{4\sqrt{2}}\left[ge^{\frac{1}{\sqrt{2}}\f}+me^{-\frac{3}{\sqrt{2}}\f}-\frac{4k}{g} e^{-\frac{1}{\sqrt{2}}\f-2\l(r)}\right]+3\Upsilon e^{\frac{1}{\sqrt{2}}\f-4\l(r)},\nn
\\
\l'e^{-f}&=-\frac{1}{4\sqrt{2}}\left[ge^{\frac{1}{\sqrt{2}}\f}+me^{-\frac{3}{\sqrt{2}}\f}+\frac{4k}{g} e^{-\frac{1}{\sqrt{2}}\f-2\l(r)}\right]-\Upsilon e^{\frac{1}{\sqrt{2}}\f-4\l(r)},
\\
\frac{\f'}{\sqrt{2}}e^{-f}&=-\frac{1}{4\sqrt{2}}\left[-ge^{\frac{1}{\sqrt{2}}\f}+3me^{-\frac{3}{\sqrt{2}}\f}+\frac{4k}{g} e^{-\frac{1}{\sqrt{2}}\f-2\l(r)}\right]+\Upsilon e^{\frac{1}{\sqrt{2}}\f-4\l(r)}.\nn
\end{align}

When we adopt new variables $x:=e^{2\l-\sqrt{2}\f}$, $F:=e^{2\sqrt{2}\f}x$, the flow equations are reduced to
\ba\label{eq:12}
\frac{dF}{dx}=\frac{F(2mgx+4k)}{x(g^{2}F-mgx)+4\sqrt{2}g\Upsilon} .
\end{align}
For a K\"ahler 4-cycle, we have an AdS$_2$ fixed point when the cycle is negatively curved $(k=-1)$ at $F=4/g^{2},\,x=2/gm$. On the other hand, for a Cayley 4-cycle, we have a supersymmetric fixed point when $k=-1,\, F=8/3g^{2},\,x=2/gm$. Series expansion solutions can be also easily worked out, and we have  
\be\label{eq:13}
F= \frac{3m}{g}x+\frac{12k}{g^{2}}+\frac{1}{g^2}\sum_{n=1}^{\infty}\frac{c_{n}}{(mgx)^{\frac{n}{2}}}.
\ee
Just like 2- and 3-cycle cases, $c_n \,(n>1)$ can be determined recursively in terms of $c_1$ when we substitute this expression into Eq.\eqref{eq:12}. Numerically we find that the flow to AdS${}_2$  corresponds to $c_1=23.538$ for K\"ahler case, and $c_1=19.7959$ for Cayley case.

\begin{figure}
    \centering
    \includegraphics[scale=0.7]{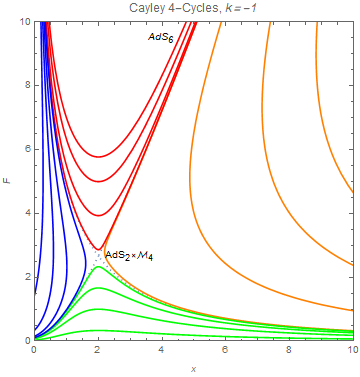}\includegraphics[scale=0.7]{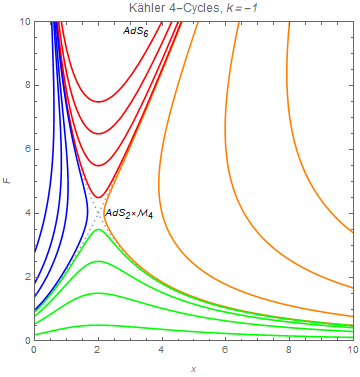}
    \caption{\label{fig3}
Flow Diagrams for negatively curved Cayley and  K\"ahler 4-cycles, $k=-1$}
\end{figure}

\begin{figure}

    \centering
    \includegraphics[scale=0.7]{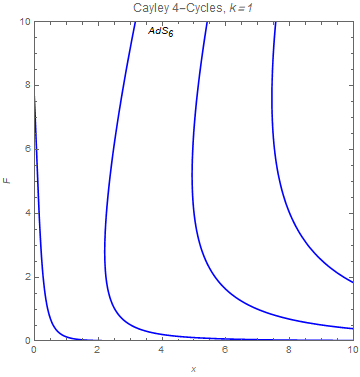}\includegraphics[scale=0.7]{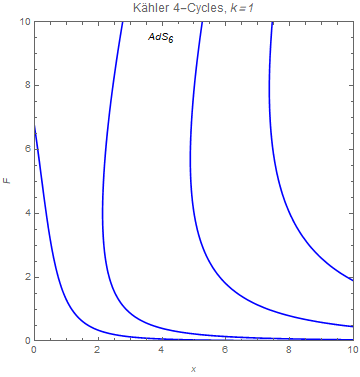}
    \caption{\label{fig4}Flow Diagrams for positively curved Cayley and  K\"ahler 4-cycles, $k=1$}
\end{figure}

Qualitatively speaking, when we analyze the UV asymptotics of $F(x)$, we notice that its behavior is similar to that of Eq.\eqref{eq:8} with substitution $d=4$. 
In IR, both good and bad type singularities exist under the criterion in \cite{Maldacena2001g}, but there is no good singularity according to the criterion in \cite{Gubser2000b}. The flow diagrams and the types of singularities with respect to corresponding limit are summarized in Fig. \ref{fig3}, Fig. \ref{fig4} and the table below.

\begin{center}
    
\begin{tabular}{|c|c|c|c|c|c|c|}
\hline
 $k$&$x$&$F$&$e^{2f}$&$\left|g_{tt}^{10d}\right|$&$V(\f)$&Type\\ \hline
  $\pm1$&$\infty$&0&0&0&$\infty$(Bad)&-
 \\
$\pm1$&0&0&$\infty$&$\infty$ &$\infty$(Bad)&Bad
\\
$\pm1$&0&Finite&$\infty$&$\infty$ &$\infty$(Bad)&Bad
 \\
$-1$&0&$\infty$&0&0&$\infty$(Bad)&-
\\\hline
\end{tabular}
\end{center}

\subsection{K\"ahler 4-Cycles as a Product of Two Riemann Surfaces}
For the K\"ahler 4-cycle case, in fact, one may consider a generalization where it is a direct product of two Riemann surfaces and allow different radii. The twisting and projection rules are 
\ba
\w^{23}+\w^{45}=g\z A^{\hat{3}},&&\half\g_{23}\e=\half\g_{45}\e=-\z T^{\hat{3}}\e.
\end{align}
Now we have four lines of BPS equations, as below. Note that they reduce to the previous BPS equations for 4-cycles Eq.\eqref{bps4c} through identification, $\l_{1}=\l_{2}$, $k_1=k_2$, and setting $\Upsilon=-\frac{1}{\sqrt{2}g^{2}m}$.
\ba
f'e^{-f}&=-\frac{1}{4\sqrt{2}}\left[ge^{\frac{1}{\sqrt{2}}\f}+me^{-\frac{3}{\sqrt{2}}\f}-\frac{2}{g} e^{-\frac{1}{\sqrt{2}}\f}(k_{1} e^{-2\l_{1}(r)}+k_{2}e^{-2\l_{2}(r)})\right]-3\Upsilon e^{\frac{1}{\sqrt{2}}\f-2\l_{1}(r)-2\l_{2}(r)},\nn
\\
\l_{1}'e^{-f}&=-\frac{1}{4\sqrt{2}}\left[ge^{\frac{1}{\sqrt{2}}\f}+me^{-\frac{3}{\sqrt{2}}\f}+\frac{2}{g} e^{-\frac{1}{\sqrt{2}}\f}(3k_{1}e^{-2\l_{1}(r)}-k_{2}e^{-2\l_{2}(r)})\right]+\Upsilon e^{\frac{1}{\sqrt{2}}\f-2\l_{1}(r)-2\l_{2}(r)},\nn
\\
\l_{2}'e^{-f}&=-\frac{1}{4\sqrt{2}}\left[ge^{\frac{1}{\sqrt{2}}\f}+me^{-\frac{3}{\sqrt{2}}\f}+\frac{2}{g} e^{-\frac{1}{\sqrt{2}}\f}(-k_{1}e^{-2\l_{1}(r)}+3k_{2}e^{-2\l_{2}(r)})\right]+\Upsilon e^{\frac{1}{\sqrt{2}}\f-2\l_{1}(r)-2\l_{2}(r)},\nn
\\
\frac{\f'}{\sqrt{2}}e^{-f}&=-\frac{1}{4\sqrt{2}}\left[-ge^{\frac{1}{\sqrt{2}}\f}+3me^{-\frac{3}{\sqrt{2}}\f}+\frac{2}{g} e^{-\frac{1}{\sqrt{2}}\f}(k_{1}e^{-2\l_{1}(r)}+k_{2}e^{-2\l_{2}(r)})\right]-\Upsilon e^{\frac{1}{\sqrt{2}}\f-2\l_{1}(r)-2\l_{2}(r)}.
\end{align}

Introducing 
 $x_{1}:=e^{2\l_{1}-\sqrt{2}\f}$, $x_{2}:=e^{2\l_{2}-\sqrt{2}\f}$, $u:=e^{2\sqrt{2}\f}x_{1}x_{2}=e^{2\l_{1}+2\l_{2}}$,
we obtain the following flow equations, where we treat $x_1,x_2$ as a function of $u$. 
\ba
\frac{dx_{1}}{du}&=\frac{x_{1}}{u}\left[\frac{g^{3}mu-g^{2}m^{2}x_{1}x_{2}+2gm(k_{1}x_{2}-k_{2}x_{1})-4}{g^{3}mu+g^{2}m^{2}x_{1}x_{2}+2gm(k_{1}x_{2}+k_{2}x_{1})-4}\right],\label{eq:22}
\\
\frac{dx_{2}}{du}&=\frac{x_{2}}{u}\left[\frac{g^{3}mu-g^{2}m^{2}x_{1}x_{2}-2gm(k_{1}x_{2}-k_{2}x_{1})-4}{g^{3}mu+g^{2}m^{2}x_{1}x_{2}+2gm(k_{1}x_{2}+k_{2}x_{1})-4}\right].\label{eq:23}
\end{align}

We find there is only one supersymmetric fixed point where $x_1=x_2$, which we already know: $
u=8/g^{3}m,\quad x_{1}=x_{2}=2/gm,\quad k_{1}=k_{2}=-1 . 
$

One may try to construct
series expansion solutions. From numerical solutions, we find that the solution aymptotes to AdS${}_6$ only if $k_1=k_2, x_1=x_2$. We thus do not present the series form of solutions here, since it should be identical with \eqref{eq:13}. For $k_{1}=-k_{2}$, we find there is a one-parameter family of numerical solutions connected to AdS$_{6}$, and its series form is as follows.
\ba
x_{1}&=\sqrt{g/(3m)}\sqrt{u}-\frac{2k_{1}}{gm}+\sum_{n=1}^{\infty}\mathcal{C}^{(1)}_{n}u^{-n/4},
\\
x_{2}&=\sqrt{g/(3m)}\sqrt{u}-\frac{2k_{2}}{gm}+\sum_{n=1}^{\infty}\mathcal{C}^{(2)}_{n}u^{-n/4},
\end{align}
where $\mathcal{C}^{(1)}_{1}=\mathcal{C}^{(2)}_{1}=\mathcal{C}_{1}$ is an integration constant, and the subleading coefficients can be found iteratively.

Below we report on the classification of IR singularities in general flows with $x_{1}\neq x_{2}$. 

\begin{center}   
\begin{tabular}{|c|c|c|c|c|c|c|}
\hline
 $x_{1}$&$x_{2}$&$F$&$e^{2f}$&$\left|g_{tt}^{10d}\right|$&$V(\f)$&Type\\ \hline
$\infty$&$0$&0&0&0&$\infty$(Bad)&-
  \\
   0&$\infty$&0&0&0&$\infty$(Bad)&-
 \\
$\infty$&$\infty$&0&0&0&$\infty$(Bad) &-
\\
0&0&0&$\infty$&$\infty$&$\infty$(Bad)&Bad
\\\hline
\end{tabular}
\end{center}

\section{Lower-Dimensional Actions and Non-Supersymmetric Fixed Points}
\label{sec:4}
It turns out that, upon application of partial twisting, $F(4)$ gauged supergravity allows various non-supersymmetric AdS solutions in addition to supersymmetric ones. They can be found either by solving the field equations in $D=6$ directly, or one can first work out a consistently truncated action in lower dimensions and look for critical points of the scalar potential thereof.

A simple approach in $D=6$ is to assume the existence of an AdS fixed point and write \cite{Naka2002}.
\ba
\label{eq:para}
e^{f}=\frac{\a}{gr}e^{-\frac{1}{\sqrt{2}}\f},&&e^{\l_{i}}=\frac{\b_{i}}{g}e^{-\frac{1}{\sqrt{2}}\f},&&\g=e^{-2\sqrt{2}\f}
,
\end{align}
where $\alpha,\beta_i,\gamma$ are constants. We, then, obtain algebraic equations involving them, and, from their solutions,
we have reproduced non-supersymmetric solutions with 2- and 3-cycles found in \cite{Naka2002} and also discovered new non-supersymmetric solutions for 4-cycles, {\it e.g.} $AdS_{2}\times\mathcal{M}^{k=1}_{\textrm{K\"ahler}}$ and $AdS_{2}\times S^{2} \times S^{2}$ fixed points. We expect they can also be obtained as near horizon geometry of $AdS_{6}$ black holes whose horizon is $\mathcal{M}^{k=1}_{\textrm{K\"ahler}}$ or $S^{2} \times S^{2}$. On the other hand, 
non-BPS AdS${}_{3}$ and AdS${}_{4}$ solutions correspond to near horizon geometry of black strings and black 2-branes, respectively.

\subsection{2- and 3- Cycles}

Let us start with the case of 2-cycles. From the field equations, we can of course double-check the supersymmetric solution with $k=-1$, 
$
\a^{2}_{BPS}=8$, $\b_{BPS}^{2}=4$, $\g_{BPS}=g/(2m).
$
There is in fact another solution which is non-supersymmetric \cite{Naka2002},
$
\a^{2}_{non-BPS}
\approx6.61921$, $
\b_{non-BPS}^{2}
\approx3.47593$, $
\g_{non-BPS}
\approx0.694146 g/m.
$

For 3-cycles, we reproduce a supersymmetric solution, $\a^{2}_{BPS}=9/2$, $\b^{2}_{BPS}=3$, $\g_{BPS}=2g/3m$, and also a non-supersymmetric one, at 
 $\a^{2}_{non-BPS}\approx5.27966$, $\b^{2}_{non-BPS}\approx3.41324$,
and $\g_{non-BPS}\approx0.507683g/m$.

\subsubsection{Lower-dimensional action for 2- and 3-cycles}
One can straightforwardly check that by keeping only the modes $\lambda,\phi$ in BPS equations discussed earlier, and allowing general metric for the $(6-d)$-dimensional part, we obtain consistently truncated lower-dimensional actions. It can be also worked out collectively for $d=2$ and $d=3$. In Einstein-frame, the result is 
\ba
\mathcal{S}^{Ein}_{6-d}&=\frac{\textrm{Vol}(\mathcal{M}_{d})}{2\k_{6}^{2}} \int d^{6-d}x \sqrt{-g_{6-d}}\Big[ \frac{1}{4}R-\frac{d}{(4-d)}\pt_{\m}\l\pt^{\m}\l -\frac{1}{2}\pt_{\m}\f\pt^{\m}\f
\nn
\\
&+\frac{kd}{4}e^{-\frac{8\l}{4-d}}+\frac{1}{8}e^{-\frac{2d\l}{4-d}}(g^2e^{\sqrt{2}\f}+4gme^{-\sqrt{2}\f}-m^2e^{-3\sqrt{2}\f}) \nn
-\frac{\t_{\mathcal{M}_{d}}}{4g^{2}}e^{-\frac{2(8-d)\l}{4-d}}e^{-\sqrt{2}\f}\Big],
\end{align}
where $\t_{\mathcal{M}_{d=2}}=2$, $\t_{\mathcal{M}_{d=3}}=3/2$. 
We record that the metric ansatz which leads to the Einstein-frame action above is 

\ba
ds^{2}_{6}=e^{-\frac{2d}{4-d}\l}ds^{2}_{6-d}+e^{2\l}ds^{2}_{\mathcal{M}_{d}} .
\end{align}

\subsubsection{Stability of Non-Supersymmetric Solutions for 2- and 3-cycles}
The stability of supersymmetric solutions is guaranteed by unbroken supersymmetry, but, for non-supersymmetric solutions, there is no such guarantee. Thus we need to work out the eigen-frequency of fluctuation modes to check the stability. In this paper, we restrict ourselves to the modes kept by $D=6$ supergravity, which are the lightest modes and intuitively most likely to lead to tachyonic modes. We consider small fluctuations of $\l$ and $\f$ around non-supersymmetric solutions of fields near non-supersymmetric AdS solutions, and  diagonalize the mass matrix for $\l$ and $\f$. 

For 2-cycles, we find 
\ba
M_{\textrm{unstable}}^{2}R^{2}\approx-3.032\leq-\frac{9}{4},&&
M_{\textrm{stable}}^{2}R^{2}\approx1.741\geq-\frac{9}{4},
\end{align}
where BF bound for AdS$_{4}$ is $M_{\textrm{scalar}}^{2}R^{2}\geq-\frac{9}{4}$, so we conclude this solution is unstable. 

For 3 cycles, we obtain  
\ba
M_{\textrm{unstable}}^{2}R^{2}\approx-1.593\leq-1,&&
M_{\textrm{stable}}^{2}R^{2}\approx-0.444\geq-1,
\end{align}
where BF bound for AdS$_{3}$ is $M_{\textrm{scalar}}^{2}R^{2}\geq-1$, so we again encounter instability. 

\subsection{4-Cycles}
One can verify the BPS solutions for negatively curved 4-cycles and also find non-BPS solutions for positively curved K\"ahler 4-cycles which are locally $S^{2}\times S^{2}$ or $\mathbb{CP}^{2}$.
\subsubsection{Fixed Point Solutions for Cayley and K\"ahler 4-Cycles}
For Cayley cycles, it turns out that there are no AdS solutions other than the BPS solution: $\a_{BPS}^{2}=2$, $\b_{BPS}^{2}=8/3$, and $\g_{BPS}=3g/(4m)$.

For K\"ahler cycles on the other hand, we find, in addition to a supersymmetric solution with $k=-1$, $\a_{BPS}^{2}=2$, $
\b_{BPS}^{2}=4$, and $\g_{BPS}=g/(2m)$, there is a non-BPS solution for $k=1$, having
\ba
\a^{2}_{non-BPS}=\frac{1}{5} \left(4-\sqrt{6}\right),&&
\b_{non-BPS}^{2}=\frac{4}{5} \left(4-\sqrt{6}\right),&&\g_{non-BPS}=\frac{1}{4} \left(2+\sqrt{6}\right)\frac{g}{m} . 
\end{align}

Although there could be solutions with different scalar curvature and radius for two Riemann surfaces, we find there is no additional fixed point than reported already in previous subsections.

\subsubsection{Two Dimensional Theories on 4-Cycles}
We here present the bosonic action for two dimensional effective theories on $\mathcal{M}_{4}$, which can be a supersymmetric four-cycle, {\it i.e.} Cayley or K\"ahler. As it is well known, one cannot move to Einstein frame through scale  transformation in 2 dimensions and that is why there is a conformal factor $e^{\l_{1}+\l_{2}}$ below. 

\ba
\mathcal{S}_{2}=&\frac{\textrm{Vol}(\mathcal{M}_{4})}{2\k^{2}_{6}}\int d^{2}x \sqrt{-g_{2}}e^{2\l_{1}+2\l_{2}}\left[\frac{1}{4}R_{2}+\frac{1}{2}(e^{-2\l_{1}}k_{1}+e^{-2\l_{2}}k_{2})\right.
\nn
\\
&+\frac{1}{2}g^{\m\n}\pt_{\m}\l_{1}\pt_{\n}\l_{1}+\frac{1}{2}g^{\m\n}\pt_{\m}\l_{2}\pt_{\n}\l_{2}+2g^{\m\n}\pt_{\m}\l_{1}\pt_{\n}\l_{2}-\frac{1}{2}\pt_{\m}\f\pt^{\m}\f
\nn
\\
&+\frac{1}{8}(g^2e^{\sqrt{2}\f}+4gme^{-\sqrt{2}\f}-m^2e^{-3\sqrt{2}\f}) \nn
\\
&\left.-\frac{\t_{\mathcal{M}_{4}}}{4g^{2}}e^{-\sqrt{2}\f}(e^{-4\l_{1}}+e^{-4\l_{2}})-\frac{\t_{\mathcal{M}_{4}}^{2}}{2m^{2}g^{4}}e^{\sqrt{2}\f-4\l_{1}-4\l_{2}}\right],
\end{align}
where $\t_{\mathcal{M}_{\text{Cayley}}}=2/3$, $\t_{\mathcal{M}_{\text{K\"ahler}}}=2$, and $\t_{\S_{1}\times\S_{2}}=2$. Note that for Cayley and K\"ahler 4-cycles as {\it e.g.} $\mathbb{C}\mathbb{P}^2$ we need to set $\l_1=\l_2$ and $k_{1}=k_{2}$. From this effective action, one can reproduce all the results above involving 4-cycles. We record the reduction ansatz for $D=6$ metric is
\ba
ds^{2}_{6}=ds^{2}_{2}+\sum_{i=1}^2 e^{2\l_{i}}ds^{2}_{\mathcal{M}_{i}}.
\end{align}

From the action and the equations of motion, we have calculated the mass eigenvalues of scalar fluctuations around the non-supersymmetric $AdS_2\times S^2\times S^2$ and the result is 
\ba
M^{2}_{1}R^{2}=\frac{3}{20} \left(6+\sqrt{6}\right),
&&
M^{2}_{2}R^{2}=3,
&&
M^{2}_{3}R^{2}=\frac{1}{4} \left(6+\sqrt{6}\right).
\end{align}
We thus find there is no unstable mode.
 
\subsection{Entropy of Black Objects in 6 Dimensions}

\subsubsection{Non-Supersymmetric $AdS_{6}$ Black Holes}

Now let us consider the Bekenstein-Hakwing entropy formula for black holes in $AdS_{6}$, with $AdS_{2}\times S^{2}\times S^{2}$ or $AdS_{2}\times \mathbb{C}\mathbb{P}^2$ as the near horizon geometry.
\be
S_{BH}=\frac{A_{H}}{4G_{N}}=\frac{2\p A_{H}}{\k_{6}^{2}},
\ee
where $G_{N}$ is six dimensional Newton constant. Horizon area with the parametrization in this section is given by
\ba
A_{H}=\frac{\prod^{2}_{i=1}\b_{i}^{2}}{g^{4}}\g\textrm{ Vol}(\mathcal{M}_{4}).
\end{align}
Therefore, entropy of non-supersymmetric $AdS_{6}$ black holes is 
\be
S^{non-BPS}_{BH}=\frac{2(3 \sqrt{6}-2)}{25g^{3}m G_{N}^{(6)}}\times\left\{\begin{array}{cc}
16\p^{2}& S^{2}\times S^{2}\textrm{ horizon}
\\
18\p^{2}& \mathbb{CP}^{2}\textrm{ horizon}
\end{array}\right.
\ee

We also confirmed the entropy of supersymmetric $AdS_{6}$ black holes concerned in \cite{Suh2018, Suh2019}. 
\ba
S^{BPS}_{BH}=\frac{2\check{s}}{g^{3}mG_{N}^{(6)}}\textrm{ Vol}(\mathcal{M}^{k=-1}_{4}),
\end{align}
where $\check{s}=2/3$, $1$ for Cayley 4-cycles and K\"ahler 4-cycles respectively. Volume of hyperbolic 4-manifolds is given in terms of Euler Characteristic, $\c$. To be specific, $\mathbb{H}_{4}$, the Bergman spaces\footnote{They are denoted by $\mathcal{M}_{B}$}, and $\mathbb{H}_{2}\times\mathbb{H}_{2}$ are used for Cayley and K\"ahler 4-cycles, and K\"ahler 4-cycles as products of two Riemann surfaces: $\textrm{ Vol}(\mathbb{H}_{4})=12\p^{2}\c(\mathbb{H}_{4})$, $\textrm{ Vol}(\mathcal{M}_{B})=6\p^{2}\c(\mathcal{M}_{B})$, and $\textrm{ Vol}(\mathbb{H}^{2}\times\mathbb{H}^{2})=16\p^{2}(\mathfrak{g}_{1}-1)(\mathfrak{g}_{2}-1)$, respectively. Note that Einstein condition, $\hat{R}_{ab}=k\d_{ab}$, is used to fix the normalization of curvature tensor of calibrated cycles in the orthonormal frame. 

With $g^{3}m=108$ for unit AdS$_{6}$ radius, one can explicitly compare with field theory result.
Holographic renormalization has been done with superpotential counter terms and boundary terms. It is natural to consider holographic renormalization of non-supersymmetric solutions which might have holographic conformal field theories as their dual theories. In contrast to the supersymmetric cases, there is no obvious choice of holographic renormalization scheme using {\it e.g.} supersymmetric counter-terms. It is thus rather far-fetched to check AdS/CFT quantitatively for non-supersymmetric examples.

\subsubsection{Supersymmetric Black 2-Branes and Strings}
(Non-)supersymmetric $AdS_{3}$ or $AdS_4$ solutions can be considered as near horizon geometry of black strings and black 2-branes. Let us recall the metric ansatz of wrapped branes on two and three cycles.

\be
ds^{2}_6=\frac{R^{2}}{r^{2}}(-dt^{2}+dr^{2}+\sum_{\a=1}^{4-d}dx^{2}_{\a})+e^{2\l}ds_{\mathcal{M}_{d}}^{2},\quad d=2,3.
\ee

In the coordinate choice above, the boundary of 
$AdS_{p+2}$ is $\mathbb{R}^{1}\times\mathbb{R}^{p}$ on which worldvolume SCFTs of black $p$-brane are defined. Since horizon and worldvolume of $p$-branes are non-compact, we used compact volume of horizon and density of entropy of black $p$-branes to obtain finite values, respectively. 

Area of horizon for effective $AdS_{4}$ and $AdS_{5}$ black holes is written in terms of $g$ and $m$.
\ba
A_{BPS}^{(4)}=\frac{2\sqrt{2}}{\sqrt{g^{3}m}}\textrm{Vol}(\mathcal{M}_{2}),&&
A_{BPS}^{(5)}=(\frac{6}{g^{3}m})^{3/4}\textrm{Vol}(\mathcal{M}_{3}).
\end{align}

The entropy density of the black strings and branes are given in 
\ba
s^{(6)}_{\textrm{BB,BPS}}=\frac{32\p}{g^{3}m\k_{6}^{2}}\textrm{Vol}(\mathcal{M}_{2}),
&&
s^{(6)}_{\textrm{BS,BPS}}=\frac{6\sqrt{6}\p}{g^{3}m\k_{6}^{2}}\textrm{Vol}(\mathcal{M}_{3}).
\end{align}

We reproduce the same entropy density formula in \cite{Bobev2017}. The special case of black strings corresponds to Brown-Henneaux central charge \cite{Brown:1986nw} in the context of AdS${}_{3}$/CFT$_{2}$. 
For supersymmetric cases, one should recall the fact that $L_{AdS_{6}}=3\sqrt{2}(3mg^{3})^{-1/4}$. 

Since non-supersymmetric solutions contain unstable scalar modes which violate BF bound, we have not considered the entropy density of non-BPS black 2-branes and strings. 

\subsubsection{Holographic Check}
For BPS solutions, we can check the holographic relations in \cite{Bobev2017} for black D2 and D1 branes with unit $AdS_{6}$ radius, i.e. $g^{3}m=108$.
\ba
s^{(6)}_{\textrm{BS,BPS}}=\frac{\sqrt{6}}{144G^{(6)}_{N}}\textrm{Vol}(\mathcal{M}_{3})=\frac{1}{6}c_{2d},
&&
s^{(6)}_{\textrm{BB,BPS}}=\frac{4\p(\mathfrak{g}-1)}{27G^{(6)}_{N}}=-\frac{4(\mathfrak{g}-1)}{9\p}\mathcal{F}_{S^{5}}=\frac{\mathcal{F}_{S^{3}\times\S_{\mathfrak{g}}}}{2\p}.
\end{align}
This agrees with the result in \cite{Bobev2017,Hosseini2018a,Crichigno2018}.
\ba
c_{2d}=-\frac{\sqrt{6}\textrm{Vol}(\mathcal{M}_{3})}{8\p^{2}} \mathcal{F}_{S^{5}},
&&
\mathcal{F}_{S^{3}}=
\mathcal{F}_{S^{3}\times\S_{\mathfrak{g}}}=-\frac{8(\mathfrak{g}-1)}{9} \mathcal{F}_{S^{5}} .
\end{align}

\section{Discussions}
\label{sec:5}
In this paper, we have analyzed all fixed points and holographic renormalization group flows associated with the geometries which describe the branes wrapping on calibrated cycles in several special holonomy manifolds with appropriate topological twists. We have also tried to determine if the IR singularities are physically admissible, but, for some cases, the Maldacena-Nu\~nez criterion and the Gubser criterion give us contradictory answers. We, thus, need to perform more elaborate analysis such as construction of black hole solutions whose horizon hides the singularity. We postpone this work to future works.

In addition, we have also worked out lower-dimensional consistently truncated action in 4, 3, and 2 dimensions. Using them, we have checked the stability of the non-supersymmetric solutions with respect to the Breitenlohner-Freedman bound. Let us emphasize that the lower dimensional actions we have presented are {\it not} the bosonic part of some supersymmetric action. We need to  consider vector and tensor fields, additionally, in the same way as  \cite{Donos2010,Colgain:2010rg,Cheung:2019pge}. One might be able to find interesting solutions such as the ones exhibiting Lifshitz-scaling \cite{Donos2010}, and we postpone this problem also to future works.

From the viewpoint of recent developments concerning the comparison using AdS/CFT, we point out that there exist gravity solutions whose field theory dual is not amenable to localization treatment. It is mainly due to insufficient amount of preserved supersymmetry. For instance, the AdS${}_2$ solution wrapped on Cayley 4-cycle has only two supercharges, and we do not know how to do the field theory side calculation. It is similar to the situation with sphere partition functions: 
${\cal N}=2$ (8 supercharges) is needed to put the theory on $S^4$ and localize \cite{Pestun:2007rz}, and similarly to put a three-dimensional theory on $S^3$ and localize one needs ${\cal N}=2$ (4 supercharges) \cite{Jafferis2012,Hama:2010av}. Our final comment is that a number of supergravity solutions are still waiting for field theory computation to catch up.

\begin{acknowledgments}
We are grateful to Hyojoong Kim for discussions and comments. MS is grateful to Francesco Benini, E. \'O Colg\'ain, P. Marcos Crichigno, Jerome P. Gauntlett, Leopoldo A. Pando Zayas, and Minwoo Suh for discussions. This work was supported by the National Research Foundation (NRF) grant 2018R1D1A1B07045414. The research of MS was also supported by a scholarship from Hyundai Motor Chung Mong-Koo Foundation.  
\end{acknowledgments}

\appendix

\section{More on $F(4)$ Gauged Supergravity\label{app:sec}}

\subsection{\label{app:subsec}Equations of Motion and Supersymmetry Transformation Rules}

The equations of motion one may derive from the above action are as follows
\ba
&R_{\m\n}=2(\pt_{\m}\f\pt_{\n}\f)-\frac{1}{8}g_{\m\n}(g^{2}e^{\sqrt{2}\f}+4gm e^{-\sqrt{2}\f}-m^{2}e^{-3\sqrt{2}\f})+2e^{-\sqrt{2}\f}(\mathcal{H}_{\m}^{\ph{m}\r}\mathcal{H}_{\n\r}-\frac{1}{8}\mathcal{H}^{\m\n}\mathcal{H}_{\m\n})\nn
\\&\quad+2e^{-\sqrt{2}\f}(F^{I\ph{a}\r}_{\m}F^{I}_{\n\r}-\frac{1}{8}g_{\m\n}F^{I}_{\a\b}F^{I\a\b})+e^{2\sqrt{2}\f}(G_{\m}^{\ph{a}\r\s}G_{\n\r\s}-\frac{1}{6}g_{\m\n}G^{\a\b\g}G_{\a\b\g}) ,
\\
&\frac{1}{\sqrt{-g}}\pt_{\m}(\sqrt{-g}g^{\m\n}\pt_{\n}\f)=-\frac{1}{4\sqrt{2}}(g^{2}e^{\sqrt{2}\f}-4gme^{-\sqrt{2}\f}+3m^{2}e^{-3\sqrt{2}\f})\nn
\\&\quad-\frac{1}{2\sqrt{2}}e^{-\sqrt{2}\f}(\mathcal{H}^{\m\n}\mathcal{H}_{\m\n}+F^{I}_{\a\b}F^{I\a\b})-\frac{1}{3\sqrt{2}}e^{2\sqrt{2}\f}G^{\m\n\r}G_{\m\n\r},
\\&\nabla_{\m}(e^{-\sqrt{2}\f}\mathcal{H}^{\m\n})=\frac{1}{6}\e^{\n\m\r\s\a\b}\mathcal{H}_{\m\r}G_{\s\a\b}, 
\\&D_{\m}(e^{-\sqrt{2}\f}F^{I\m\n})=\frac{1}{6}\e^{\n\m\r\s\a\b}\mathcal{F}^{I}_{\m\r}G_{\s\a\b},
\\&\nabla_{\r}(e^{2\sqrt{2}\f}G^{\r\m\n})=\frac{1}{4}\e^{\m\n\r\s\a\b}F^{I}_{\r\s}F^{I}_{\a\b}+me^{-\sqrt{2}\f}\mathcal{H}^{\m\n},
\end{align}
where $SU(2)$ covariant derivative is $DF^{I}=dF^{I}+g\e^{I}_{\ph{I}JK}A^{J}\we F^{K}$.

The sign of $U(1)$ gauge connection ${\mathcal{A}}$ inside covariant derivative is a choice of convention. For all the solutions we consider in this paper, $\mathcal{A}=0$.

For the local supersymmetry transformation parameter we use symplectic Majorana spinors, $\e^{i}$, which have a symplectic index, $i=1,2$. There is an isomorphism between $USp(2)$ and $SU(2)$. We identify the symplectic index with $SU(2)$ index \cite{Romans1986d}. 

There are fermionic fields including gravitini and gaugini. Supersymmetry transformation rules of these fermions are given here. 
\ba
\d\ps_{\m{}i}=&\pt_{\m}\e_{i}+\frac{1}{4}\w_{\m\n\r}\g^{\n\r}\e_{i}+gA^{\hat{I}}_{\m}(T^{\hat{I}})_{i}^{\ph{i}j}\e_{j}+\frac{i}{8\sqrt{2}}(ge^{\frac{\f}{\sqrt{2}}}+me^{-3\frac{\f}{\sqrt{2}}})\g_{\m}\g_{7}\e_{i}\nn
\\&-\frac{i}{4\sqrt{2}}(\g_{\m}^{\ph{a}\n\r}-6\d_{\m}^{\ph{a}\n}\g^{\r})e^{-\frac{\f}{\sqrt{2}}}\g_{7}F^{\hat{I}}_{\n\r}T^{\hat{I}\ph{i}j}_{i}\e_{j}\nn
\\&+\frac{i}{8\sqrt{2}}e^{-\frac{\f}{\sqrt{2}}}\mathcal{H}_{\n\r}(\g_{\m}^{\ph{a}\n\r}-6\d_{\m}^{\ph{a}\n}\g^{\r})\e_{i}-\frac{1}{24}e^{\sqrt{2}\f}G_{\n\r\s}\g_{7}\g^{\n\r\s}\g_{\m}\e_{i},
\\
\d\c_{i}=&\frac{i}{\sqrt{2}}\g^{\m}\pt_{\m}\f\e_{i}-\frac{1}{4\sqrt{2}}(ge^{\frac{\f}{\sqrt{2}}}-3me^{-3\frac{\f}{\sqrt{2}}})\g_{7}\e_{i}\nn
\\
&+\frac{1}{2\sqrt{2}}e^{-\frac{\f}{\sqrt{2}}}\g^{\n\r}\g_{7}F^{\hat{I}}_{\n\r}T^{\hat{I}\ph{i}j}_{i}\e_{j}-\frac{1}{4\sqrt{2}}e^{-\frac{\f}{\sqrt{2}}}\mathcal{H}_{\n\r}\g^{\n\r}\e_{i}+\frac{i}{12}G_{\m\n\r}\g_{7}\g^{\m\n\r}\e_{i}.
\end{align}

The spinor $\epsilon_i$ is chiral, and chosen to satisfy the below projection condition.
\beq
i \g_{1}\g_{7}\e_{i}=\e_{i},&&\e_{i}=e^{f/2}\e_{i,\textrm{ const}},
\eeq
where $\g_{1}=e_{1}^{\ph{1}r}\g_{r}$, and the chirality matrix is defined in the orthonormal frame, $\g_{7}=\g_{0}\g_{1}\g_{2}\g_{3}\g_{4}\g_{5}$.

\subsection{The Supersymmetric AdS$_{6}$ Vacuum and the Killing Spinors}

With constant dilaton, the equations of motion and Killing spinor equations are reduced as below. 

\ba
R_{\m\n}&=-\frac{1}{8}g_{\m\n}(g^{2}e^{\sqrt{2}\f}+4gme^{-\sqrt{2}\f}-m^{2}e^{-3\sqrt{2}\f}),
\\
\d\ps_{\m{}i}&=\pt_{\m}\e_{i}+\frac{1}{4}\w_{\m\n\r}\g^{\n\r}\e_{i}+\frac{i}{8\sqrt{2}}(ge^{\frac{\f}{\sqrt{2}}}+me^{-3\frac{\f}{\sqrt{2}}})\g_{\m}\g_{7}\e_{i}=0,
\\
0&=-\frac{g^{2}}{4\sqrt{2}}e^{\sqrt{2}\f}(1-\frac{m}{g}e^{-2\sqrt{2}\f})(1-3\frac{m}{g}e^{-2\sqrt{2}\f}),
\\
\d\c_{i}&=-\frac{g}{4\sqrt{2}}e^{\frac{\f}{\sqrt{2}}}(1-3\frac{m}{g}e^{-2\sqrt{2}\f})\g_{7}\e_{i}=0.
\end{align}

From the last two equations, we obtain a solution for dilaton field: $e^{-2\sqrt{2}\f}=g/(3m)$. This solution is one of extrema of $V(\f)$.  The well-known choice of vanishing dilaton, $g=3m$, is a convenient choice for uplifting to $D=10$ type \Rom{2} supergravities. 

The Einstein equations and a relation between $AdS_{6}$ radius and Ricci scalar give the below equations for $f$ with in the metric ansatz adopted.
\ba
-5f''e^{-2f}&=-\frac{5}{18}g^{2}e^{\sqrt{2}\f},
\\
(f''+4(f')^{2})e^{-2f}&=\frac{5}{18}g^{2}e^{\sqrt{2}\f},
\\
(2(f')^{2}+f'')e^{-2f}&=\frac{3}{18}g^{2}e^{\sqrt{2}\f}=\frac{3}{L_{AdS_{6}}^{2}}.
\end{align}
By solving the equations, we obtain $AdS_{6}$ in Poincar\'e coordinates.
\be
ds^{2}_6=\frac{L_{AdS_{6}}^{2}}{r^{2}}(-dt^{2}+dr^{2}+\sum_{\a=1}^{4}dx^{2}_{\a})
\ee
Dilaton and the $AdS_{6}$ radius are written in terms of $g$ and $m$.
\ba
e^{-2\sqrt{2}\f}=\frac{g}{3m},&& L_{AdS_{6}}^{4}=3^{4}\cdot2^{2}\frac{1}{3mg^{3}}.
\end{align} 
The left equation is the Killing spinor equation for gravitini in the following form. 
\ba
\nabla_{\m}\e_{i}&=-\frac{i}{2L_{AdS_{6}}}\g_{\m}\g_{7}\e_{i}
\end{align}
Solutions of this type of Killing spinor equations are well-known.

\section{Details on the UV Expansion}
In the coordinates convention we adopt, $r\rightarrow0$ limit corresponds to asymptotic $AdS_{6}$ in UV. Due to singularities from the denominators in $dx_{i}/dF$, we used UV expansion to analyze the BPS equations. General UV ans\"atze in terms of $r$ are
\ba{\label{eqnUVr}}
e^{f}=\frac{L_{AdS_{6}}}{r}+\sum_{k=0}^{\infty}\a_{k}\frac{r^{k}}{L^{k}_{AdS_{6}}},
&&
e^{\l_{i}}=\frac{L_{AdS_{6}}}{gr}+\sum_{k=0}^{\infty}\e^{(i)}_{k}\frac{r^{k}}{L^{k}_{AdS_{6}}},
&&
e^{\f/\sqrt{2}}=\frac{3\sqrt{2}}{gL_{AdS_{6}}}+\sum_{k=0}^{\infty}\g_{k}\frac{r^{k}}{L^{k}_{AdS_{6}}}.
\end{align}

Here, employing $x$ and $F$ is more convenient to analyze instead of $r$. Integration constants from the UV expansions are identified for the flow from UV to the fixed points. Except for the 4-cycle consist of 2 Riemann surfaces, UV expansion series are written in a unified manner in terms of $x$. 
\ba
\frac{dF}{dx}=\frac{F(4kx+2mgx^{2})}{x[x(g^{2}F-mgx+(4-d)k)+4\sqrt{2}g\Upsilon]},&&F=3\frac{m}{g}x+\frac{3dk}{g^{2}}+\sum_{n=1}^{\infty}\mathcal{C}^{(F)}_{n}x^{-\frac{n}{2}}.
\end{align}
Instanton densities for each cases:
\ba
\Upsilon_{d=2,3}=0,&&\Upsilon_{\textrm{Cayley}}=-\frac{1}{3\sqrt{2}g^{2}m},&&\Upsilon_{\textrm{K\"ahler}}=-\frac{1}{\sqrt{2}g^{2}m}.
\end{align}
All of the expansion coefficients are written in terms of $g$, $m$, and $\mathcal{C}^{(F)}_{1}$.
\ba\left.
\begin{array}{ccc}
\mathcal{C}^{(F)}_{2}=12(\frac{\sqrt{2}}{g}\Upsilon-\frac{d}{mg^{3}}),
&
\mathcal{C}^{(F)}_{3}=-\frac{(6+d)k}{2mg}\mathcal{C}^{(F)}_{1},
&
\mathcal{C}^{(F)}_{4,\ph{.}d=4}=\frac{-\frac{1}{2}g^{2}\mathcal{C}^{(F)}_{1}\mathcal{C}^{(F)}_{1}-16k\mathcal{C}^{(F)}_{2}}{3mg},\\[0.5 em]
\end{array}\right.
\\
\mathcal{C}^{(F)}_{n+2}=-\frac{2(2n+dn+4)k\mathcal{C}^{(F)}_{n}+4\sqrt{2}(n-2)g\Upsilon \mathcal{C}^{(F)}_{n-2}+\sum_{a=1}^{n-1}ag^{2}\mathcal{C}^{(F)}_{a}\mathcal{C}^{(F)}_{n-a}}{2(1+n)mg},&&(n\geq n^{*}),\nn
\end{align}
where $n^{*}=2$ for $d=2,3$ and $n^{*}=3$ for $d=4$.

For the case of two Riemann surfaces, we obtained ten recursion relations from the flow equations. For $n\geq2$, recursion relations are given as
\ba
0&=g^{3}m\frac{(n+4)}{2}\mathcal{C}^{(1)}_{n}-2n\mathcal{C}^{(1)}_{n-4}
+g^{2}m^{2}\sum_{a=-4}^{n-2}\sum_{b=-2}^{a+2}\frac{(n-a-8)}{2}\mathcal{C}^{(1)}_{n-4-a}\mathcal{C}^{(1)}_{a-b}\mathcal{C}^{(2)}_{b}
\nn
\\
&+gm\sum_{a=-2}^{n-2}\{k_{1}(n-a)\mathcal{C}^{(2)}_{a}+k_{2}(n-8-a)\mathcal{C}^{(1)}_{a}\}\mathcal{C}^{(1)}_{n-a},
\\
0&=(1)\leftrightarrow(2).\nn
\end{align}
To match with \eqref{eqnUVr}, one can check the asymptotic $x$ in terms of $r$,
\ba
x_{\textrm{UV}}=\frac{6}{mg^{3}}\frac{1}{r^{2}}.
\end{align}

\section{Lower Dimensional Metric Ansatz and Dimensional Analysis}

Right after the dimensional reduction with the metric ansatz we employ, we obtain string frame actions. For three and four dimensional theories, one can use a conformal transformation to go to Einstein frame. The right choice can be easily worked out, 
\ba
g^{Ein}_{\m\n}=e^{\frac{2d}{4-d}\l}g^{Str}_{\m\n}.
\end{align}
The mass spectrum of fluctuation modes should be calculated in Einstein frame for $d<4$.

\section{Equations of Motion in Lower Dimensions}

Lower dimensional effective theories are parametrized by their dimensionality $6-d$ and $\t_{\mathcal{M}_{d}}$ determined by the ansatz for $SU(2)$ gauge fields. Equations of motion of effective theories are given here\footnote{Note that the equations in this appendix are given in a manipulated form to facilitate the comparison with 6d equations after variation.}. Using the ansatz, one can recover the 6d equations of motion for each case.  
\subsection{Equations of Motion in 3, 4 Dimensions}

Equations of motion for $d=2,3$ cycles can be obtained using variational principle. Here is the Einstein equation valid for 3 or 4 spacetime dimensions.
\ba
R_{\a\b}&=-g_{\a\b}\left[\frac{d}{(4-d)}\frac{1}{\sqrt{-g_{6-d}}}\pt_{\m}(\sqrt{g_{6-d}}\pt^{\m}\l)+\frac{1}{8}e^{-\frac{2d\l}{4-d}}(g^2e^{\sqrt{2}\f}+4gme^{-\sqrt{2}\f}-m^2e^{-3\sqrt{2}\f})\right. \nn
\\
&\left.+\frac{1}{4}e^{-\frac{2d\l}{4-d}}e^{-\sqrt{2}\f}(\t_{\mathcal{M}_{d}}\frac{e^{-4\l}}{g^{2}})\right]+\frac{4d}{(4-d)}\pt_{\a}\l\pt_{\b}\l +2\pt_{\a}\f\pt_{\b}\f  
\end{align}
This is the equation for $\l$:
\ba
\frac{1}{\sqrt{-g_{6-d}}}\pt_{\m}(\sqrt{g_{6-d}}\pt^{\m}\l)&=ke^{-\frac{8\l}{4-d}}+\frac{1}{8}e^{-\frac{2d\l}{4-d}}(g^2e^{\sqrt{2}\f}+4gme^{-\sqrt{2}\f}-m^2e^{-3\sqrt{2}\f}) \nn
\\
&-(8-d)\frac{\t_{\mathcal{M}_{d}}}{4dg^{2}}e^{-\frac{2(8-d)}{4-d}\l}e^{-\sqrt{2}\f}
\end{align}
The equation for scalar $\f$:
\ba
\frac{1}{\sqrt{-g_{6-d}}}\pt_{\m}(\sqrt{-g_{6-d}}\pt^{\m}\f)
&=-\frac{\sqrt{2}}{8}e^{-\frac{2d\l}{4-d}}(g^2e^{\sqrt{2}\f}-4gme^{-\sqrt{2}\f}+3m^{2}e^{-3\sqrt{2}\f})\nn
\\
&-\frac{\sqrt{2}\t_{\mathcal{M}_{d}}}{4g^{2}}e^{-\frac{2(8-d)}{4-d}\l}e^{-\sqrt{2}\f}
\end{align}
With the ansatz, the following equations of the fixed points for 2, 3-cycles are written in terms of (\ref{eq:para}).
\ba
\left.\begin{array}{l}
\a^{2}=\frac{3\b^{4}}{(2-k\b^{2})}\ph{abcdef}\textrm{(2-cycles)}
\\
 2\frac{m}{g}\g=-k\frac{6}{\b^{2}}+8\b^{-4}-1 
\\
0=\frac{4}{\b^{4}}-k\frac{96}{\b^{6}}+\frac{64}{\b^{8}}+k\frac{28}{\b^{2}}+5
\end{array}\right.,&&
\left.\begin{array}{l}
\a^{2}=\frac{2\b^{4}}{(1-\b^{2}k)}\ph{abcdefg}\textrm{(3-cycles)}
\\
\frac{m}{g}\g=3\frac{1}{\a^{2}}-\frac{3}{2\b^{2}}k-\half
\\
0=\frac{9}{\b^{8}}-36\frac{k}{\b^{6}}+\frac{26}{\b^{4}}+\frac{28}{\b^{2}}k+5
\end{array}\right.
\end{align}

By solving the equation for $\b^2$ with $k=-1$, one would obtained two physical solutions for BPS and non-BPS fixed points. 
 
\subsection{Equations of Motion in 2 Dimensions}

Two dimensional Einstein equations in string frame are given in relatively more complicated form. 
\ba
R_{\m\n}&-\frac{2}{\sqrt{-g}}g_{\n\b}\pt_{\m}(\sqrt{-g}g^{b \b}\pt_{b}(\l_{1}+\l_{2}))\nn
\\
&-2g_{\a\m}(\pt_{\d}g^{\d\a})\pt_{\n}(\l_{1}+\l_{2})+2g_{\a\m}g_{\n\b}\pt_{\g}(g^{\a \b})\pt_{\d}(\l_{1}+\l_{2})g^{\g\d}\nn
\\
&-\pt_{\m}(\l_{1}+\l_{2})g^{\a\b}(\pt_{\a} g_{\b\n}+\pt_{\b} g_{\a\n}-\pt_{\n}g_{\a\b})+\pt_{\b}(\l_{1}+\l_{2})g^{\a\b}\pt_{\a} g_{\m\n}\nn
\\
=&+g_{\m\n}[-\half\pt_{\b}(\l_{1}+\l_{2})\pt_{\a}g^{\a\b}+\half g_{\a\b}g^{\g\d}\pt_{\d}(\l_{1}+\l_{2})\pt_{\g}g^{\a \b}\nn
\\
&-\frac{g^{2}}{8}(e^{\sqrt{2}\f}+4\frac{m}{g}e^{-\sqrt{2}\f}-\frac{m^2}{g^2}e^{-3\sqrt{2}\f})-\frac{\t_{\mathcal{M}_{4}}}{4} e^{-\sqrt{2}\f}(\frac{e^{-4\l_{1}}+e^{-4\l_{2}}}{g^{2}})+\frac{\t_{\mathcal{M}_{4}}^{2}}{2m^{2}g^{4}}e^{\sqrt{2}\f-4\l_{1}-4\l_{2}}\nn
\\
&-\frac{1}{4}g^{\g\d}\pt_{\g}(\l_{1}+\l_{2})g^{\a\b}(\pt_{\a} g_{\b \d}+\pt_{\b} g_{\a\d}-\pt_{\d}g_{\a\b})+\frac{1}{4}\pt_{\b}(\l_{1}+\l_{2})g^{\a\b}g^{\g\d}\pt_{\a} g_{\g\d}]\nn
\\
&+2\pt_{\m}\l_{1}\pt_{\n}\l_{1}+2\pt_{\m}\l_{2}\pt_{\n}\l_{2}+2\pt_{\m}\f\pt_{\n}\f-g_{\m\n}\frac{2\t_{\mathcal{M}_{4}}^{2}}{m^{2}g^{4}}e^{\sqrt{2}\f+4f-4\l_{1}-4\l_{2}}.
\end{align}

We have three equations of motion for scalars: 
\ba
\frac{1}{\sqrt{-g}}\pt_{\m}(\sqrt{-g}g^{\m\n}\pt_{\n}\l_{1})&=-2g^{\m\n}\pt_{\m}(\l_{1}+\l_{2})\pt_{\n}\l_{1}+e^{-2\l_{1}}k_{1}\nn
\\
&+\frac{1}{4}g^{\g\d}\pt_{\g}(\l_{1}+\l_{2})g^{\a\b}(\pt_{\a} g_{\b\d}+\pt_{\b} g_{\a\d})+\half\pt_{\b}(\l_{1}+\l_{2})\pt_{\a}g^{\a\b}\nn
\\
&+\frac{g^{2}}{8}(e^{\sqrt{2}\f}+4\frac{m}{g}e^{-\sqrt{2}\f}-\frac{m^2}{g^2}e^{-3\sqrt{2}\f})-\frac{\t^{2}_{\mathcal{M}_{4}}}{2m^{2}g^{4}}e^{\frac{2}{\sqrt{2}}\f-4(\l_{1}+\l_{2})}\nn
\\
&-\frac{\t_{\mathcal{M}_{4}}}{4}e^{-\sqrt{2}\f}(\frac{3e^{-4\l_{1}}-e^{-4\l_{2}}}{g^{2}}),
\\
\frac{1}{\sqrt{-g}}\pt_{\m}(\sqrt{-g}g^{\m\n}\pt_{\n}\l_{2})
&=-2g^{\m\n}\pt_{\m}(\l_{1}+\l_{2})\pt_{\n}\l_{2}+e^{-2\l_{2}}k_{2}\nn
\\
&+\frac{1}{4}g^{\g\d}\pt_{\g}(\l_{1}+\l_{2})g^{\a\b}(\pt_{\a} g_{\b\d}+\pt_{\b} g_{\a\d})+\half\pt_{\b}(\l_{1}+\l_{2})\pt_{\a}g^{\a\b}\nn
\\
&+\frac{g^{2}}{8}(e^{\sqrt{2}\f}+4\frac{m}{g}e^{-\sqrt{2}\f}-\frac{m^2}{g^2}e^{-3\sqrt{2}\f})-\frac{\t^{2}_{\mathcal{M}_{4}}}{2m^{2}g^{4}}e^{\frac{2}{\sqrt{2}}\f-4(\l_{1}+\l_{2})}\nn
\\
&-\frac{\t_{\mathcal{M}_{4}}}{4}e^{-\sqrt{2}\f}(\frac{-e^{-4\l_{1}}+3e^{-4\l_{2}}}{g^{2}}),
\\
\frac{1}{\sqrt{-g}}\pt_{\m}(\sqrt{-g}g^{\m\n}\pt_{\n}\f)=&-2\pt_{\m}(\l_{1}+\l_{2})\pt^{\m}\f-\frac{\sqrt{2}g^{2}}{8}(e^{\sqrt{2}\f}-4\frac{m}{g}e^{-\sqrt{2}\f}+3\frac{m^2}{g^2}e^{-3\sqrt{2}\f}) \nn
\\
&\frac{\sqrt{2}\t^{2}_{\mathcal{M}_{4}}}{2m^{2}g^{4}}e^{\frac{2}{\sqrt{2}}\f-4(\l_{1}+\l_{2})}-\frac{\sqrt{2}\t_{\mathcal{M}_{4}}}{4}e^{-\sqrt{2}\f}(\frac{e^{-4\l_{1}}+e^{-4\l_{2}}}{g^{2}}).
\end{align}

With (\ref{eq:para}) and the ans\"atze, equations are reduced to 4 algebraic equations for two Riemann surfaces.
\ba
\left.\begin{array}{c}
\frac{8}{\a^{2}}=(1+4\frac{m}{g}\g-\frac{m^{2}}{g^{2}}\g^{2})+\frac{48}{\frac{m^{2}}{g^{2}}\g^{2} \b_{1}^{4}\b_{2}^{4}}+4(\frac{1}{\b_{1}^{4}}+\frac{1}{\b_{2}^{4}}),
\\
8\frac{k_{1}}{\b_{1}^{2}}=-(1+4\frac{m}{g}\g-\frac{m^{2}}{g^{2}}\g^{2})+\frac{16}{\frac{m^{2}}{g^{2}}\g^{2} \b_{1}^{4}\b_{2}^{4}}+4(\frac{3}{\b_{1}^{4}}-\frac{1}{\b_{2}^{4}}),
\\
8\frac{k_{2}}{\b_{2}^{2}}=-(1+4\frac{m}{g}\g-\frac{m^{2}}{g^{2}}\g^{2})+\frac{16}{\frac{m^{2}}{g^{2}}\g^{2} \b_{1}^{4}\b_{2}^{4}}+4(-\frac{1}{\b_{1}^{4}}+\frac{3}{\b_{2}^{4}}),
\\
0=-(1-4\frac{m}{g}\g+3\frac{m^{2}}{g^{2}}\g^{2})+\frac{16}{\frac{m^{2}}{g^{2}}\g^{2}\b_{1}^{4}\b_{2}^{4}}-4(\frac{1}{\b_{1}^{4}}+\frac{1}{\b_{2}^{4}}).\end{array}\right.
\end{align}
For single 4-cycles, a set of algebraic equations is given in the following way. 
\ba
\left.\begin{array}{l}
\frac{1}{\a^{2}}=\frac{8}{9\frac{m^{2}}{g^{2}}\g^{2}}\frac{1}{\b^{8}}+\frac{2}{3\b^{4}}-\frac{k}{\b^{2}}\ph{ab}\textrm{(Cayley)}
\\
0=1+2\frac{m}{g}\g-\frac{4}{3\b^{4}}+6\frac{k}{\b^{2}}-\frac{16}{9\b^{8}\frac{m^{2}}{g^{2}}\g^{2}}
\\
\frac{m}{g}\g=1\pm\frac{\sqrt{3 \b^{4}+6k\b^{2}-4}}{\sqrt{3} \b^{2}}
\end{array}\right.,
&&
\left.\begin{array}{l}
\frac{1}{\a^{2}}=\frac{8}{\frac{m^{2}}{g^{2}}\g^{2}}\frac{1}{\b^{8}}+\frac{2}{\b^{4}}-\frac{k}{\b^{2}}\ph{ab}\textrm{(K\"ahler)}
\\
\frac{m}{g}\g=1\pm\frac{1}{\b^{2}}\sqrt{\b^{4}-4+2k\b^{2}}
\\
0=1+2\frac{m}{g}\g-\frac{4}{\b^{4}}+6\frac{k}{\b^{2}}-\frac{16}{\b^{8}\frac{m^{2}}{g^{2}}\g^{2}}
\end{array}\right.
\end{align}

From last two equations, we obtained algebraic equation of $\b^{2}$ for Cayley and K\"ahler 4-cycles respectively.
\ba
0&=1-\frac{4}{3 \b^4}+\frac{6 k}{\b^2}+2 (1\pm \frac{1}{\b^2}\sqrt{\b^4+2k \b^2-4/3})-\frac{16}{9 \b^8(1\pm\frac{1}{\b^{2}}\sqrt{\b^4+2k \b^2-4/3})^2}
\end{align}
For Cayley 4-cycles, $-$ sign should be taken for the physical solution, $\b^{2}_{BPS}=8/3$ with $k=-1$.
\ba
0=1+2(1\pm\frac{1}{\b^{2}}\sqrt{\b^{4}-4+2k\b^{2}})-\frac{4}{\b^{4}}+6\frac{k}{\b^{2}}-\frac{16}{\b^{8}(1\pm\frac{1}{\b^{2}}\sqrt{\b^{4}-4+2k\b^{2}})^{2}}
\end{align}
For K\"ahler 4-cycles, both of the BPS and non-BPS solutions are obtained in the case of $k=-1$. However, we also confirmed non-BPS solution exists for $k=1$

\bibliography{ref.bib}

%
\end{document}